\documentclass[twocolumn,english,aps,prl,reprint,groupedaddress,notitlepage,nobibnotes,nofootinbib,preprintnumbers,showpacs]{revtex4-1}
\pdfoutput=1
\usepackage{lmodern}

\usepackage[T1]{fontenc}
\usepackage[latin9]{inputenc}
\usepackage{geometry}
\usepackage{color}
\usepackage{babel}
\usepackage{amsmath}
\usepackage{amssymb}
\usepackage{graphicx}
\usepackage{esint}
\usepackage[unicode=true,pdfusetitle,
 bookmarks=true,bookmarksnumbered=false,bookmarksopen=false,
 breaklinks=false,pdfborder={0 0 1},backref=false,colorlinks=true]
 {hyperref}
\hypersetup{
 citecolor=blue,linkcolor=blue,urlcolor=blue}
\makeatletter

 \@ifundefined{textcolor}{}
 {
   \definecolor{BLACK}{gray}{0}
   \definecolor{WHITE}{gray}{1}
   \definecolor{RED}{rgb}{1,0,0}
   \definecolor{GREEN}{rgb}{0,1,0}
   \definecolor{BLUE}{rgb}{0,0,1}
   \definecolor{CYAN}{cmyk}{1,0,0,0}
   \definecolor{MAGENTA}{cmyk}{0,1,0,0}
   \definecolor{YELLOW}{cmyk}{0,0,1,0}
 }
\usepackage{babel}
\makeatletter
\def\simgt{\mathrel{\lower2.5pt\vbox{\lineskip=0pt\baselineskip=0pt
           \hbox{$>$}\hbox{$\sim$}}}}
\def\simlt{\mathrel{\lower2.5pt\vbox{\lineskip=0pt\baselineskip=0pt
           \hbox{$<$}\hbox{$\sim$}}}}
\makeatother

\newcommand{\be}{\begin{equation}}
\newcommand{\ee}{\end{equation}}
\newcommand{\bea}{\begin{eqnarray}}
\newcommand{\eea}{\end{eqnarray}}
\newcommand{\Ref}[1]{Ref.~\cite{#1}}
\newcommand{\Fig}[1]{Fig.~\ref{#1}}
\newcommand{\Eq}[1]{Eq.~(\ref{#1})}

\newcommand{\mPl}{m_{\rm Pl}}

\begin{document}

\preprint{\hbox{CALT-68-2879} }

\title{Naturalness and the Weak Gravity Conjecture}

\author{Clifford Cheung}
\author{Grant N. Remmen}
\affiliation{Walter Burke Institute for Theoretical Physics \\California Institute of Technology, Pasadena, CA 91125}
\date{\today}
\email{clifford.cheung@caltech.edu}\email{gremmen@theory.caltech.edu}
\begin{abstract}
The weak gravity conjecture (WGC) is an ultraviolet consistency condition asserting that an Abelian force requires a state of charge $q$ and mass $m$ with $q>m/m_{\rm Pl}$. We generalize the WGC to product gauge groups and study its tension with the naturalness principle for a charged scalar coupled to gravity. Reconciling naturalness with the WGC either requires a Higgs phase or a low cutoff at $\Lambda \sim q m_{\rm Pl}$.  If neither applies, one can construct simple models that forbid a natural electroweak scale and whose observation would rule out the naturalness principle.
\end{abstract}

\maketitle

\section{Introduction\label{intro}}

The naturalness principle asserts that operators not protected by symmetry are unstable to quantum corrections induced at the cutoff.  As a tenet of effective field theory, naturalness has provided a key motivation for new physics at the electroweak scale.  However, the discovery of the Higgs boson \cite{ATLAS,CMS} together with null results from direct searches has led many to revisit naturalness as a fundamental principle.  Rather than amend naturalness to fit the data, we  instead explore its interplay with established concepts in quantum field theory.  

Our focus will be the weak gravity conjecture (WGC) \cite{WGC}, which states that a consistent theory of gravity coupled to an Abelian gauge theory must contain a state of charge $q$ and mass $m$ satisfying\footnote{We define the Planck mass, $m_{\rm Pl}$, such that \Eq{eq:WGC} is saturated for an extremal black hole.}
\bea
q & > & m/\mPl,
\label{eq:WGC}
\eea
i.e., gravity is the weakest force.  While \Eq{eq:WGC} is certainly true of electromagnetism, \Ref{WGC} convincingly argued that it is a universal consistency condition of all healthy quantum field theories.

However, in theories with fundamental scalars \Eq{eq:WGC} runs afoul of naturalness because it bounds a quadratically divergent mass by a logarithmically divergent charge.  For small charge, \Eq{eq:WGC} forbids a natural spectrum in which scalars have masses near the cutoff.
We illustrate this contradiction with scalar quantum electrodynamics (QED) coupled to general relativity, but this tension is a ubiquitous feature of any model with a hierarchy problem and a small charge.  We also generalize \Eq{eq:WGC} to the case of multiple forces and particles.

 As we will show, reconciling naturalness with \Eq{eq:WGC} requires a revision of the original theory: either the gauge symmetry is spontaneously broken or new degrees of freedom enter prematurely at the cutoff
\bea
\Lambda &\sim& q m_{\rm Pl}.
\label{eq:cutoff}
\eea
\Ref{WGC} conjectured \Eq{eq:cutoff} with the stronger interpretation that $\Lambda$ signals the complete breakdown of four-dimensional quantum field theory.  Supporting this claim with compelling string theoretic examples, \Ref{WGC} fell short of a general argument.  However, if one asserts the primacy of naturalness, then our logic provides a reason from quantum field theory for new states at $\Lambda$.

To illustrate these ideas we present simple, concrete extensions of the standard model (SM) in which a natural value of the electroweak scale---at the Planck scale---is incompatible with \Eq{eq:WGC} due to a new millicharged force.  These models offer the unique opportunity to test naturalness experimentally.  Indeed, either naturalness reigns, in which case \Eq{eq:cutoff} demands a low cutoff, or it fails.  
Absent additional ultralight states, a discovery of this millicharged force would then invalidate naturalness and mandate an unnatural electroweak scale.  In particular, \Eq{eq:WGC} would disallow a natural electroweak scale and the hierarchy problem would arise from as-yet-unknown ultraviolet dynamics.  More generally, a fifth force discovery of any kind would invalidate the interpretation of $\Lambda$ advocated in \Ref{WGC} as the cutoff of four-dimensional quantum field theory.  If, as conjectured in \Ref{WGC}, this breakdown is a universal feature of all string compactifications, such an observation would also falsify string theory.

\section{Evidence for the WGC\label{evidence}}
Let us summarize the justification for the WGC \cite{WGC}.  
Consider a $U(1)$ gauge theory with charged species labeled by $i$,  each representing a particle (anti-particle) of charge $q_i$ ($-q_i$) and mass $m_i$.  We define dimensionless charge-to-mass ratios,
\bea
z_i &=&   q_i \mPl/m_i,
\eea
so \Eq{eq:WGC} implies that there exists some particle $i$ with $z_i > 1$.
The authors of \Ref{WGC} offered theoretical evidence in support of \Eq{eq:WGC}.  They presented many examples from field theory and string theory, all satisfying \Eq{eq:WGC}.   Further, they argued that \Eq{eq:WGC} reconciles the inherent inconsistency of exact global symmetries with the na\"ively innocuous $q\rightarrow 0$ limit of a gauge theory.   This limit yields an exact global symmetry; however, such charges are not conserved by quantum gravity \cite{BekensteinNoHair1,BekensteinNoHair2} because, in accordance with no-hair theorems \cite{HawkingEllis}, a stationary black hole is fully characterized by its mass, spin, and charge.   

Of course, examples and consistency with no-hair theorems only provide circumstantial evidence for \Eq{eq:WGC}.   Importantly, \Ref{WGC} also argues for \Eq{eq:WGC} via {\it reductio ad absurdum}, drawing only on general relativity, conservation of charge and energy, and  minimal assumptions about the ultimate theory of quantum gravity.  Consider a black hole of charge $Q$ and mass $M$ decaying solely to particles of species $i$, which can occur via Hawking radiation or Schwinger pair production \cite{QuantumHair,BlackHoleSchwinger}.
By charge conservation, $Q/q_i$ particles are produced.  Conservation of energy dictates that the total rest mass of the final state, $m_i Q /q_i$, be less than $M$.  In terms of the black hole charge-to-mass ratio, $Z =  Q \,  \mPl /M$, this implies $z_i >Z$. An extremal black hole corresponds to $Z=1$ and is stable unless some state $i$ exists for which $z_i >1$.   If \Eq{eq:WGC} fails, the spectrum contains a large number of stable black hole remnants, in tension with holographic bounds  \cite{tHooft,Bousso} and afflicted with various quantum gravitational and thermodynamic pathologies  \cite{SusskindRemnants,GiddingsRemnant}.

\section{The Limits of Naturalness\label{naturalness}}

The WGC is straightforward at tree-level, but radiative corrections introduce subtleties.  In fermionic QED, $q$ and $m$ run with renormalization scale, as does their ratio, na\"ively making $q/m$ ambiguous; however, as \Ref{WGC} notes, the appropriate scale to evaluate $q/m$ is the physical mass of the particle.  This is the mass scale that is relevant to the kinematics of extremal black hole decay, which provides the justification for the WGC.

However, the radiative stability question becomes more interesting in scalar QED:
\bea
{\cal L} &=&- \frac{1}{4}  F_{\mu\nu }^2 +  |D_\mu \phi|^2 - m^2 |\phi|^2 - \frac{\lambda}{4}|\phi|^4,
\label{eq:sQED}
\eea
where $D_\mu = \partial_\mu + i q A_\mu$ is the gauge covariant derivative. As for any effective field theory, we assume an ultraviolet cutoff, $\Lambda$, above which new physics enters.
Since $\phi$ is a fundamental scalar, its mass is radiatively unstable and corrected by $m^2  \rightarrow  m^2 + \delta m^2$ where
\bea
\delta m^2 &=& \frac{ \Lambda^2}{16\pi^2}(a q^2 + b \lambda).
\eea
Here  $a$ and $b$ are dimensionless numerical coefficients.  We assume that $\delta m^2$ is positive so that the theory remains in the Coulomb phase.
In a natural theory, the physical mass of $\phi$ cannot be parametrically smaller than its radiative corrections.  Equivalently, the counterterm for the scalar mass should not introduce a delicate cancellation.  This is formally equivalent to requiring that the coefficients $a$ and $b$ take on ${\cal O}(1)$ values.

Let us set the physical mass squared for $\phi$ to its natural value, $\delta m^2$, which the WGC forbids from exceeding its charge in Planck units.  The charge-to-mass ratio of $\phi$ is
\bea
z &=& \frac{4 \pi \mPl}{\Lambda} \frac{1}{\sqrt{a+ b \lambda /q^2}},
\label{eq:zdef}
\eea
where the WGC implies that $z>1$.
If $q^2 \gg \lambda$, then
\bea
\Lambda &<& \frac{4 \pi \mPl }{\sqrt{a}},
\label{eq:WGCweak}
\eea
which is the reasonable requirement that the cutoff not exceed the Planck scale.  

Turning to the opposite hierarchy, $q^2 \ll \lambda$, which is also radiatively stable, we find that the WGC implies
\bea
\Lambda &<& 4 \pi \mPl \sqrt{\frac{q^2 }{b \lambda }}.
\label{eq:WGCstrong}
\eea
As $q^2 /\lambda \rightarrow 0$, a sensible cutoff requires $b\rightarrow 0$, indicating mandatory fine-tuning in order to satisfy the WGC.  We are left with a remarkable conclusion: scalar QED with $ q^2 \ll \lambda$ and natural masses fails \Eq{eq:WGC} and is thus inconsistent with a quantum theory of gravity.  

We have not traded a mass scale hierarchy problem for an equivalent hierarchy problem of couplings.  Small charges are radiatively stable and thus technically natural.  In principle, $q^2 \ll \lambda$ is no worse than the small electron Yukawa coupling. 

To reconcile naturalness with \Eq{eq:WGC}, one alternative is to argue that the original theory---scalar QED with $q^2 \ll \lambda$---is impossible.  For example, this would be true if nature does not permit fundamental scalars or if a hierarchy among couplings is somehow strictly forbidden.  However, there are far less drastic options, elaborated below, if one modifies the original scalar QED theory.

\medskip

\noindent $\textit{i})$ {\bf Radiative corrections induce the Higgs phase.} It is possible that quantum effects generate a tachyon for  $\phi$, Higgsing the theory.  Charge becomes ill-defined; the charge and mass eigenbases need not commute, leaving $q/m$ ambiguous. Further, the WGC is not justified in the Higgs phase.  The original argument for the WGC \cite{WGC} relied on stable extremal black holes.  However, no-hair theorems imply that there are no stationary black hole solutions supporting classical hair from a massive photon \cite{NoHairHiggs}, independent of the size of the black hole relative to the Higgs scale.  If a black hole accretes a massive-$U(1)$-charged particle, it briefly supports an associated electric field, but after a time of order the photon Compton wavelength, it balds \cite{balding} when the gauge field is radiated away to infinity or through the horizon. 

\medskip

\noindent $\textit{ii})$ {\bf New physics enters below the Planck scale.} The simplest way to reconcile the WGC with naturalness is for the effective field theory to break down at a cutoff defined by \Eq{eq:WGCstrong}.  There could be new light states regulating quadratic divergences of $\phi$, effectively lowering $\Lambda$.  This option resolves the contradiction tautologically by eliminating the hierarchy problem altogether.  However, a more interesting alternative occurs when the new states do not couple to $\phi$.  The quadratic divergence of $\phi$ is robust and $m$ is large.  If one of these new states satisfies \Eq{eq:WGC}, then $\phi$ is irrelevant: the WGC and naturalness are reconciled.  Thus, 
asserting naturalness offers \Eq{eq:WGCstrong} as a more precise version of the low cutoff conjecture of \Ref{WGC} stated in \Eq{eq:cutoff}.

\section{ More Forces, More Particles\label{forces}}

Extending our results to various charged species of different spins, the WGC implies that at least one state in the spectrum must satisfy  \Eq{eq:WGC} after taking into account radiative corrections. Naturalness is violated in parameter regions with a hierarchy between charges and couplings that generate quadratic divergences (quartic couplings, Yukawa couplings).  

The story becomes more interesting for product gauge symmetries.  Consider a gauge group $ \prod_{a=1}^N U(1)_a $ and particles $i$ with charges $q_{ia}$ and masses $m_i$.  We represent the charges,  $\vec q_i =q_{ia}$, and charge-to-mass ratios, $\vec z_i  = q_{ia} \mPl /m_i$, as vectors of $SO(N)$, the symmetry transforming the $N$ photons among each other.  If present, photon kinetic mixing can be removed by a general linear transformation on the photons, which is equivalent to redefining charge vectors of states in the theory. 

To generalize the WGC for multi-charged particles, \Eq{eq:WGC} is inadequate and requires upgrading to a constraint on $\vec q_i$ and $m_i$. \Ref{WGC} briefly alluded to this scenario, but detailed analysis will reveal quantitative differences between the WGC as applied to a single $U(1)$ versus many.   By symmetry, the proper generalized WGC must be $SO(N)$ invariant.
Na\"ively, the WGC could require at least one species $i$ with $|\vec z_i| > 1$.  However, this is insufficient---it guarantees the existence of one particle of large total charge, but preserves stability for orthogonally-charged extremal black holes.
 A stricter alternative is that for each $U(1)$ there exists a species $i$ charged under that $U(1)$ with $|\vec z_i|>1$.  Curiously, this is still actually weaker than the true generalized WGC.

To determine the proper generalized WGC, we revisit black hole decay kinematics.  Consider a black hole of charge $\vec Q$, mass $M$, and charge-to-mass ratio $\vec Z =  \vec Q  \, \mPl/ M$ decaying to a final state comprised of $n_i$ particles of species $i$. Charge and energy conservation imply 
$\vec Q = \sum_i n_i \vec q_i $ and
$M > \sum_i n_i m_i$.
If $\sigma_i = n_i m_i /M$ is the species $i$ fraction of the total final state mass, then $\vec Z = \sum_i \sigma_i \vec z_i $ and $
1 > \sum_i \sigma_i$; decay requires that $\vec Z$ be a subunitary weighted average of $\vec z_i$.  This criterion has a geometric interpretation in charge space.  Draw the vectors $\pm \vec z_i$ corresponding to the charge-to-mass ratio of each fundamental particle in the spectrum.  A weighted average of $\vec z_i$ defines the convex hull spanned by the vectors, delineating the space of $\vec Z$ that is unstable to decay.  Any state outside the convex hull is stable. Since extremal black holes correspond to $|\vec Z| =1$, the generalized WGC requires that the convex hull spanned by $\pm \vec z_i$ contain the unit ball.   

  \begin{figure}[t]
\begin{center}
\includegraphics[width=.47\textwidth]{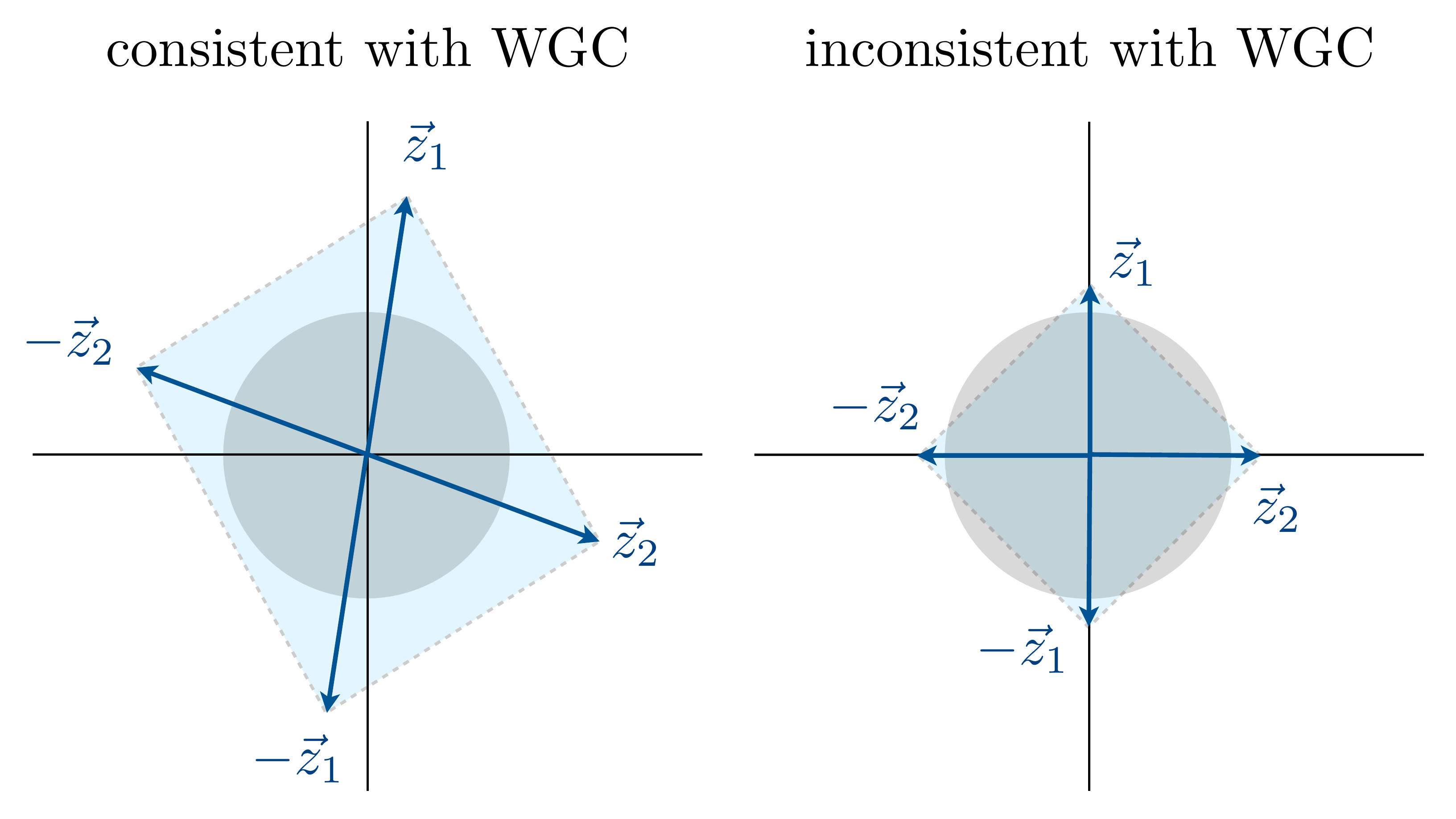}
\end{center}
\vspace*{-0.75cm}
\caption{Vectors representing charge-to-mass ratios for two species charged under two Abelian gauge symmetries.   When the convex hull defined by these vectors contains the unit ball, then extremal black holes can decay to particles and the condition of the WGC is satisfied.    }
\label{hull} 
\end{figure}

Consider a model of two Abelian factors and two charged states. The left and right panels of \Fig{hull} represent two possible choices for the charge-to-mass ratios of the particles.  Black holes of all possible charges are represented by the unit disc.  The left panel of \Fig{hull} depicts a theory that is consistent with the WGC: the unit disc is contained in the convex hull.  Extremal black holes, the boundary of this disc, can decay.
However, the right panel of \Fig{hull} depicts a theory that violates the WGC: there are regions of the unit disc not within the convex hull, corresponding to stable black hole remnants.  Remarkably, this theory fails the WGC despite the fact that $|\vec z_1| >1$ and $|\vec z_2|>1$.  Simple geometry shows that the WGC imposes the more stringent constraint:
\bea
(\vec z_1^{\; 2}-1)(\vec z_2^{\; 2}-1) > (1+ |\vec z_1 \cdot \vec z_2|)^2.
\label{eq:truebound}
\eea 
For example, given orthogonal charges of equal magnitude, $|\vec z_1| = |\vec z_2| = z$ and $\vec z_1 \perp \vec z_2$, \Eq{eq:truebound} implies $z > \sqrt{2}$, manifestly stronger than the $z>1$ condition required for theories with a single $U(1)$.
Note that the WGC places constraints on $\vec z_1$ and $\vec z_2$ that are not mathematically independent.  Were a particular value of $\vec z_1$ experimentally observed, this would fix a bound $\vec z_2^{\; 2} > (1- 1/\vec z_1^{\; 2})^{-1}$.

A similar analysis can be applied for $N$ Abelian factors and $N$ charged states.  Suppose each particle is charged under a single $U(1)$, with equal magnitude charge-to-mass ratios, so $z_{ia} = \delta_{ia} z$ for some $z$.  The convex hull defined by $\pm \vec z_i$ is an $N$-dimensional cross-polytope of circumradius $z$.  The largest ball contained in the cross-polytope has radius $z /\sqrt{N}$.  Requiring that the radius of this ball be greater than unity then implies $z > \sqrt{N}$, parametrically stronger than the condition required for a single Abelian factor.  

The WGC constraint grows at large $N$ for fixed physical Planck scale $m_{\rm Pl}$.  However, the presence of $N$ additional species generally renormalizes the strength of gravity \cite{Dvali, NimaLandscape,Nflation} as $\delta m_{\rm Pl}^2 \sim N \Lambda^2 / 16\pi^2$.  If corrections enhance $m_{\rm Pl}$  by a factor of $\sqrt{N}$, all factors of $N$ encountered in our earlier analyses cancel.  That is, in a theory with fixed Lagrangian parameters and cutoff, the limit from the WGC is $N$-independent at large $N$.  A similar phenomenon was discussed in Ref.~\cite{WGC} for $N$ Abelian factors Higgsed to a $U(1)$ subgroup.  The large-$N$ limit introduces a $Z_2^N$ symmetry, which is subject to the large-$N$ species bounds considered in Ref.~\cite{Dvali}.  

The multi-charge generalized WGC has implications for naturalness.  Consider a $U(1)^N$ gauge theory with scalars $\phi_i$ of charges $\vec q_i$ and masses $m_i$, 
\bea
{\cal L} &=& - \frac{1}{4} \sum_a F_{\mu\nu a}^2 + \sum_i |D_\mu\phi_i|^2 - m_i^2 |\phi_i|^2 - \frac{\lambda_i }{4} |\phi_i|^4, \qquad
\label{eq:scalarQEDs}
\eea
where $D_\mu \phi_i = (\partial_\mu + i \sum_a q_{ia} A_{\mu a})\phi_i$.   Radiative corrections send $m_i^2 \rightarrow m_i^2 + \delta m_i^2$, where
\bea
\delta m_i^2 &=& \frac{ \Lambda^2}{16\pi^2} (a_i {\vec q}_i^{\,2} + b_i \lambda_i)
\eea
and $a_i$ and $b_i$ are ${\cal O}(1)$ ultraviolet-sensitive coefficients.
The charge-to-mass ratio vector for $\phi_i$ is
\bea
\vec z_i &=& \frac{4 \pi \mPl}{\Lambda} \frac{ \vec q_i }{ |\vec q_i|   }\frac{1}{\sqrt{a_i+ b_i \lambda_i /\vec q_i^{\;2}}}.
\eea
A necessary albeit insufficient condition for the WGC is that, for each $U(1)$, there is a state $i$ charged under that Abelian factor such that $|\vec z_i|>1$.  This implies
\bea
\Lambda &<&4 \pi \mPl \times \left\{ 
 \begin{array}{cc}\displaystyle \frac{1 }{\sqrt{a_i}} &, \quad  \vec q_i^{\; 2} \gg \lambda_i  \\ \\
\displaystyle \sqrt{ \frac{\vec q_i^{\;2}}{b_i \lambda_i } } &, \quad  \vec q_i^{\; 2} \ll \lambda_i
 \end{array}\right.  .
\label{eq:radcorr}
\eea
As for the single Abelian case, $\vec q_i^{\; 2} \gg \lambda_i$ corresponds to the reasonable requirement of a sub-Planckian cutoff, while $ \vec q_i^{\; 2} \ll \lambda_i$ implies tension with naturalness. 
However, the most stringent requirement of the WGC---that the convex hull spanned by $\pm \vec z_i$ contain the unit ball---places a stronger limit than \Eq{eq:radcorr} by a factor of order $\sqrt{N}$ for fixed $\mPl$.

\section{ The Hierarchy Problem}
\label{hierarchy}

 We have presented explicit models in which naturalness contradicts \Eq{eq:WGC}.  We now construct theories in which natural values of the electroweak scale---at the cutoff---are similarly incompatible.  In these models, strict adherence to naturalness implies either a Higgs phase or a parametrically low cutoff given by \Eq{eq:cutoff}.
 
 The obvious path is to relate the electroweak scale to the mass $m$ of a particle that carries a tiny charge $q$.  The SM gauge couplings are ${\cal O}(1)$, so we require an additional $U(1)$ gauge symmetry beyond the SM.  It is tempting to charge the Higgs, but this will spontaneously break the $U(1)$, invalidating the applicability of the WGC.

However, we can charge the SM fermions under a very weakly gauged unbroken $U(1)_{B-L}$ symmetry.  Current limits on $U(1)_{B-L}$ require $q\lesssim 10^{-24}$ \cite{EotWash1,EotWash2} and will likely be improved by several orders of magnitude by astrophysical \cite{TripleSystem}, lunar ranging \cite{Lunar}, and satellite-based \cite{Step, Galileo, Microscope} tests of apparent equivalence principle violation.  
To cancel anomalies we introduce a right-handed neutrino $\nu_{\rm R}$ that combines with the left-handed neutrino $\nu_{\rm L}$ to form a $U(1)_{B-L}$ preserving Dirac mass term of the form $m_\nu  \bar{\nu}_{\rm L} \nu_{\rm R} + {\rm h.c.}$, where $m_\nu \sim y_\nu v$ is controlled by the electroweak symmetry breaking scale. The particle with the largest charge-to-mass ratio is the lightest neutrino.  Assuming its mass is of order the neutrino mass scale, $m_\nu  \lesssim $ 0.1 eV \cite{Neutrino1,Neutrino2}, we fix the charge to 
a technically natural albeit tiny value: $q \sim m_\nu / \mPl \sim 10^{-29}$.
For this value of $q$, \Eq{eq:WGC} is just marginally satisfied by the lightest neutrino.
While such a charge is permitted in quantum field theory, it may be difficult to engineer in string theory if $q$ arises from a string coupling constant requiring dilaton stabilization at large field values.  Similar issues arise in theories of large extra dimensions and it is a detailed question of string moduli stabilization whether this is possible.
In any case,
at fixed Yukawa coupling $y_\nu$, were the electroweak scale any higher than its measured value, \Eq{eq:WGC} would fail.  In this model, regions of parameter space favored by naturalness---and an electroweak scale at the cutoff---are inconsistent with \Eq{eq:WGC}.
Strictly speaking, this logic hinges on the absence of additional $U(1)_{B-L}$ charged states lighter than the neutrino.  Depending on the cosmological history, however, such particles may be constrained experimentally by primordial nucleosynthesis.

Our model offers a direct experimental test of naturalness by virtue of a very specific prediction: a new gauge boson very weakly coupled to the SM.  
As discussed earlier, the assumption of naturalness mandates either a Higgs phase or a low cutoff.    The discovery of a fifth force would rule out the former, while current sensitivities would for the latter imply $\Lambda \lesssim $ keV from \Eq{eq:cutoff}.  In the absence of such ultralight states, the observation of a $U(1)_{B-L}$ gauge boson at $q \sim 10^{-29}$ would then simultaneously falsify the naturalness principle and suggest an ultraviolet-dependent reason for the why the weak scale takes an unnatural value.   Moreover, given present sensitivities, a fifth force discovery of any kind would falsify string theory to the extent to which it predicts the strong interpretation of  $\Lambda$ as the scale at which four-dimensional quantum field theory breaks down \cite{WGC}.

This mechanism can generally be incorporated into any theory where the electroweak scale sources the mass of a $U(1)$ millicharged state, e.g., dark matter \cite{HiggsDescendants} charged under an unbroken $U(1)$ dark force.  For weak scale dark matter, a charge of $q \sim 10^{-16}$ is sufficient to saturate \Eq{eq:WGC}.

\medskip
\smallskip

\noindent {\it Acknowledgments}: We thank Nima Arkani-Hamed, Sean Carroll, Meimei Dong, Stefan Leichenauer, Jesse Thaler, and Mark Wise for useful discussions and comments.  We also thank the referees for helpful observations on this work.  C.C.~is supported by a DOE Early Career Award under Grant No.~DE-SC0010255.  G.N.R.~is supported by a Hertz Graduate Fellowship and a NSF Graduate Research Fellowship under Grant No.~DGE-1144469.

 \bibliographystyle{apsrevGNR}
 
\bibliography{HierarchyBibliography}

\end{document}